# Gammaless GRB's?

C Sivaram and Kenath Arun

Indian Institute of Astrophysics, Bangalore

**Abstract:** The Compactness Problem in GRB's has been resolved by invoking the Lorentz factors associated with the relativistic bulk motion. This scenario applies to GRB's where sufficient energy is converted to accelerate the ejected matter to relativistic speeds. In some situations this may not be a possible mechanism. Hence the gamma rays are trapped in the region. Here we look at these possible scenarios, where the neutrino pair production process dominates, and the neutrinos are able to escape freely. This could give a scenario where release of neutrinos precedes the gamma ray emission, which is much attenuated, possibly explaining why fewer GRB's are observed than what is expected.

Gamma ray bursts (GRB's) have now become a regular observational feature in current astronomy. Observations of afterglows revealed that these titanic releases of gamma ray energies occur at large distances of ~Gpc. This implied source energies released of well over $10^{44} J$ in gamma rays! [1]

The relativistic energies acquired by the ejected material helped resolve what was earlier known as the compactness problem. The optical depth of the source (with $F$ being the observed fluence in gamma rays) is given by:

$$\tau = \frac{f_{Phot} \sigma_T F D^2}{m_e c^2 R_S^2} \qquad \ldots (1)$$

For duration $\Delta t$ of the burst, the source size $R_S \leq c\Delta t$, $\sigma_T$ is the Thompson cross section, $D$ is the distance of the burst and $f_{Phot}$ is the fraction of photons which can produce gamma rays in the pair annihilation process: $\gamma \leftrightarrow e^- + e^+$.

For typical values ($F = 10^{-6} ergs/cm^2$, $D \approx 1 Gpc$, $\Delta t \sim 1s$), typical optical depths are very huge, that is, $\tau \approx 10^{15} - 10^{16}$! So gamma rays cannot escape from this region, i.e. from the area around the central engine. [2]

Then how are the gamma rays seen? This is the compactness problem. The resolution of the problem lies in considering the Lorentz factors associated with the relativistic bulk motion, with $\gamma \sim 100$. This would give: [2]

$$R_S = \gamma^2 c \Delta t \qquad \ldots (2)$$

There is additional $\gamma^{2\alpha}$ in the spectral index ($\gamma^{\alpha}$ for each photon, $\alpha = 2$ or 3).

This gives an overall contribution of $\gamma^{4+2\alpha}$ in the modified formula for the optical depth, that is: [3]

$$\tau = \frac{f_{Phot} \sigma_T F D^2}{\gamma^{2\alpha} m_e c^2 (\gamma^2 c \Delta t)^2} = \frac{f_{Phot} \sigma_T F D^2}{\gamma^{4+2\alpha} m_e c^2 (c \Delta t)^2} \qquad \ldots (3)$$

For $\alpha = 2$, to have an optical depth $\leq 1$, $\gamma \approx 10^2 (v = 0.995c)$, $\gamma^{4+2\alpha} \sim 10^{16}$ and for $\alpha = 3$, $\gamma \approx 50$.

The above picture applies to GRBs where sufficient energy is converted to accelerate the ejected matter to relativistic speeds. It is also commonly thought that this may be essentially the difference between supernovae (SN) and GRB's, in the sense, that the latter involve matter ejected at relativistic energies (the total rest energy corresponding to that of the collapse of a massive star, perhaps a WR star). There have been associations of GRB's with type Ic SN, whose progenitors are WR stars.

For short duration GRB's, the popular model is merger of two compact objects, i.e. neutron stars (NS) or tidal break up of a NS by a black hole. It is possible that in some situations the matter is not accelerated to relativistic speeds and the gamma rays are indeed trapped inside the region. In this case the optical depth is very high (as given by equation (1)).

However in this case, as the temperatures could be $\sim 10^{10} K$ (corresponding to $2 m_e c^2 / k_B$), we have the neutrino pair annihilation process dominating: [4]

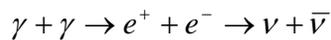

The reaction rate and energy released goes as $T^9$, where $T$ is the temperature. Essentially energy released per unit volume is given by:

$$\varepsilon \sim n_{e^+} n_{e^-} \sigma_W V k_B T \qquad \ldots (4)$$

Both $n_{e^+}$ and $n_{e^-}$ are proportional to $T^3$, $\sigma_W$, the weak cross section goes as $T^2(E^2)$ and energy of each neutrino is $\sim k_B T$. So $\varepsilon$ scales as $T^9$.

$$\varepsilon \propto T^9 \qquad \ldots (5)$$

Indeed we can write:

$$\varepsilon \sim \frac{G_F^2 E^2}{(\hbar c)^4} n_{e^+} n_{e^-} k_B T \qquad \ldots (6)$$

Where $G_F = 1.5 \times 10^{-49}\, erg\, cm^3$ is the universal Fermi weak interaction constant.

$$n_{e^+} \approx n_{e^-} \sim \left(\frac{k_B T}{\hbar c}\right)^3 \qquad \ldots (7)$$

$(E \sim k_B T)$

This gives $\varepsilon \approx 10^{20}\, ergs\, cm^{-3}$. For a volume $(c\Delta t)^3$, where $\Delta t \sim 1s$, this implies that $10^{44} J$ of neutrino energy is produced in a second. So about $10^{58}$ neutrinos are released (with average energy of several MeV).

This implies that we could have many gamma ray bursts where no gamma rays are produced, but only neutrinos. With the neutrino cross section, $\sigma_W$, equation (1) gives,

$$\tau \approx 0.1 \qquad \ldots (8)$$

So neutrinos should be freely able to escape. There is no associated compactness problem in this case! So there could be many gamma ray bursts which do not produce gamma rays, but only high energy neutrinos!

If the source of the central engine is merger of two neutron stars (like for the popular model of short duration GRB's), then most of the binding energy is anyway expected to be released through neutrinos $(\sim 10^{53}\, ergs)$ in a duration of few seconds. As neutrinos couple weakly (to matter) the radiation force may not accelerate the ejected material to relativistic speeds. The conversion $\nu + \bar{\nu} \rightarrow \gamma + \gamma$ can take place, which would fall off with distance R, from the central source as $1/R^8$, so that a 'pressure gradient' is set up.

Again as the magnetic fields in this picture of the central engines (i.e. neutron star merger) are expected to be large $(\sim 10^{12} G)$, there could be associated gamma-neutrino processes such as plasma-neutrino losses, especially for neutrinos with a magnetic moment (i.e. $\gamma \rightarrow \nu\bar{\nu}$). The emission rate density is given by: [5, 6, 7]

$$\dot{E}_\nu = \frac{\mu_\nu^2 \omega_0^4}{c^3} \int_{\omega_0}^{\infty} \frac{\omega(\omega^2 - \omega_0^2) d\omega}{\exp(\hbar\omega/k_B T) - 1} \quad \ldots (9)$$

Where, $\mu_\nu$ is the neutrino magnetic moment and $\omega_0$ is the plasma frequency.

Again neutrino Bremsstrahlung processes scale as $\sim T^6$. Net result of the above processes is to lower the gamma ray flux and enhance the neutrino flux considerably. Upper limit on the neutrino luminosity of such sources from phase-space considerations have been given in ref. [8].

Again as the neutrinos drain away energy from the source region, the optical depth (see equation (1)) could drop steeply (even without relativistic motion!) and gamma ray emission could subsequently follow but with much reduced intensity. The afterglow would now correspond more to that of a typical SN. Considering the wide variety of possible phenomena, scenarios such as the one above could also be kept in mind while discussing extremely energetic events.

It may explain (apart from beaming factor) why fewer gamma ray bursts are seen than what is expected. However in the case of neutron star mergers, gravitational waves would be detectable, as it is independent of the optical depth. So the signature of such gammaless GRB's could be simultaneous detection of neutrinos and gravitational waves.

**Concluding remarks:** In this article we discuss a possible scenario where sufficient energy of the GRB is not converted to accelerate the ejected matter to relativistic speeds and hence the gamma rays are trapped in the region. Here, the neutrino pair production process dominates, and the neutrinos are able to escape freely. In this case, the much attenuated gamma ray emission is preceded by the release of neutrinos. This could possibly be the reason why fewer GRB's are observed than what is expected. Since gravitational waves are not affected by the optical depth, its release will accompany the neutrino emission, in the case of mergers of neutron stars.